\documentstyle[prl,aps,epsfig]{revtex}
\begin{document}
\title{Holographic Discreteness of Inflationary Perturbations}
\author{Craig J. Hogan}
\address{Astronomy and Physics Departments, 
University of Washington,
Seattle, Washington 98195-1580}
\maketitle
\begin{abstract}
 The  holographic entropy bound is used to estimate the  quantum-gravitational discreteness of inflationary perturbations. 
In the context of    scalar
inflaton perturbations produced during standard slow-roll inflation,  but assuming that 
horizon-scale perturbations ``freeze out'' in discrete steps   separated by one bit of
total observable entropy, it is shown that  the Hilbert space of
  a typical horizon-scale inflaton perturbation is equivalent to that of  about $10^5$ binary spins---
approximately  the inverse of the final scalar metric perturbation amplitude, independent of other parameters.
 Holography thus suggests that in a broad  class of fundamental theories, inflationary perturbations carry  a  limited amount of
information (about $10^5$ bits per mode) and should therefore  display discreteness not predicted by the standard field theory.
Some manifestations  of
this discreteness may be observable  in cosmic background anisotropy.
\end{abstract}

\section{Introduction}

The origin of cosmological perturbations now appears to be well understood  from  the quantum
theory of fields in curved spacetime  
\cite{Starobinsky:1979ty,Hawking:1982cz,Guth:1982ec,Bardeen:1983qw,Starobinsky:1982ee,Halliwell:1985eu,Grishchuk:1993ds}. 
They originate during inflation as zero point fluctuations of the quantum modes of various
fields--- the inflaton   giving rise to scalar perturbations, the graviton to tensor
perturbations.
 The
 field quanta in the original fluctuations convert to classical perturbations as they
pass through the de Sitter-like event horizon;   they are then parametrically  amplified
by an  exponential factor during the many subsequent e-foldings of inflation, creating an
enormous number of coherent quanta in phase with the original quantum seed perturbation.
From the classical point of view,  the quantum fluctuations create
  perturbations in the classical gravitational gauge-invariant potential  $\phi_m$\cite{bardeen1980},
leading to  observable background anisotropy and  large scale structure\cite{cobeDMR,bennett,gorski,boom,pryke,halverson,max,bond01}. 
All stages of this process are under good calculational control, even the conversion of quantum
to classical
regimes
\cite{Grishchuk:1989ss,Grishchuk:1990bj,Albrecht:1994kf,Lesgourgues:1997jc,Polarski:1996jg,kiefer}.
The phase and amplitude of the large-scale classical perturbation modes observed today are a direct result of the quantum field activity
during inflation; indeed, the pattern of microwave anisotropy on the largest scales corresponds to a
faithfully amplified image  of microscopic field configurations as they froze out during inflation. Roughly speaking, each
hot or cold patch on the  sky derives originally from about one quantum.

The standard  calculation  of these processes\cite{Mukhanov:1992me,lythriotto} uses a
semiclassical  approximation: spacetime is assumed to be classical (not quantized), and the
perturbed fields (the inflaton and graviton) are described using relativistic quantum field
theory, essentially (in the limit of  free massless fields) an infinite collection of quantized
harmonic oscillators. The Hilbert space of this system is infinite, so although the fields are
quantized, they are continuously variable functions that can assume any values. The creation of
the particles can be viewed as an effect of the nonadiabatic expansion, and the ``collapse of the
wavefunction'' (in this case, ``freezing out'') can be described as a unitary quantum process of state squeezing. The theory
generically predicts random-phase gaussian noise with a continuous spectrum determined by the parameters of the
inflaton potential.  In this approximation, there is no telltale  signature
 of    quantum discreteness in the  observable classical
remnant--- the anisotropy of the background radiation.
 Although sky maps contain images of ``single quanta,'' their spectrum is continuous and the amount of information  is in principle
infinite.

It has always been acknowledged that this description is incomplete, and will be
modified by including a proper account of spacetime quantization.  
  Although the fundamental theory of such quantization is not known, a ``holographic entropy bound'' already  constrains with remarkable
precision the total number of fundamental quantum degrees of freedom.
  The complete Hilbert space of a bounded volume is
  finite and discrete  rather than infinite and continuous, limiting
the range of accessible configurations in any region to a definite, calculable number.   In particular this limit
applies to   inflaton quanta   collapsing into  classical metric perturbations.   This paper  uses the holographic
entropy bound
to  estimate   where   field theory breaks down in the inflationary analysis,   
    the effective dimension of the Hilbert space for the observable perturbations when they freeze out, 
the maximum amount of information contained in the perturbations, and
the level at which quantum-gravitational discreteness  appears in cosmic background
anisotropy. The main result is that in fundamental theories where the holographic entropy bound arises from discrete fundamental
 eigenstates,
  the amount of information in the anisotropy is remarkably limited: it can be  described with only about $10^5$
bits per sky-harmonic mode, implying that the  perturbations
should be pixelated in some way. In principle, this effect may be observable, and provide concrete data on the discrete elements or
eigenstates of 
 quantum gravity.

\section{Holographic Bound on  Information Content}

Quantitative insights into the
Hilbert space of inflationary spacetimes come from the  
thermodynamics   of   black
 holes
\cite{Bekenstein:1972tm,Bardeen:gs,Bekenstein:1973ur,Bekenstein:1974ax,Hawking:1975sw,Hawking:1976ra,Bekenstein:1981jp,bekenstein02} 
and   de Sitter spacetimes \cite{Gibbons:1977mu}.
 The black hole entropy derived from thermodynamic reasoning appears to be complete, in the sense that it
includes all possible degrees of freedom of mass-energy that make up the hole. If we insist that
black holes are quantum mechanical objects that obey unitary evolution, they must somehow encode all the
information   counted in this entropy, then radiate it to
spatial infinity as Hawking quanta as they evaporate 
\cite{Banks:1984by,'tHooft:1985re,Susskind:1993if,stephens}. Such considerations led   't Hooft
and Susskind 
 to  propose \cite{thooft93,susskind95}
a  ``holographic principle''\cite{Bigatti:1999dp} for  physical systems:
   the total entropy $S$
   within any surface is  bounded by one quarter  of the area $A$ of the surface in
Planck units. (Unless otherwise indicated, throughout this paper we adopt $\hbar=c=G=1$, with Planck mass
$m_{Planck}=(\hbar c/G)^{1/2}=1.22\times 10^{19}$ GeV.) This is an
``absolute'' entropy; the dimension
$N
$ of the Hilbert space  is given by $e^S$, and the total number of distinguishable quantum states 
available to the system is given by a binary number with $n=S/\ln 2=A/4\ln 2$
 digits \cite{bekenstein01}. 
 
Bousso\cite{bousso02} has collected and reviewed these arguments, and proposed a   more rigorous 
formulation called the ``covariant entropy bound'':   the area of a surface gives a bound on total entropy, not of the
enclosed spacelike 3-volume  but of the  3-volume defined by certain  lightsheets propagated from the surface. This bound is
apparently universal in nature; at least, no counterexamples to it have been found. Although the bound seems to hold in all
physical situations we can imagine, there does not seem to be any way to derive why this should be so, given only the physics we have.
The fact that it always works presumably reflects a deep structure in quantum gravity.

Although  a true  derivation of
the covariant entropy
bound is not yet known, its   seems to originate from a  
fundamental theory that incorporates discrete elements or eigenstates. Recent examples of theories
that display this property explicitly include loop quantum gravity\cite{ashtekar02} (where
discrete eigenstates appear at a very early stage and appear to be a
fundamental feature) and M-theory, where it is demonstrated  in particular situations where discrete
degrees of freedom can be explicitly traced via holography (such as 
AdS5/CfT) to discrete symmetries of projective, dual field theories.  These
types of fundamental structures would also impose  a   logarithmically finite constraint    on the number  of possible
solutions of fields during inflation--- that is,  jumps between possible configurations
correspond to changes  ``in the exponent'' of the number of states, in the
same way that adding just one more discrete element (e.g., a spin-1/2 particle)  to a system of
$n$ binary spins discontinuously grows the Hilbert space  from dimension $2^n$ to
$2^{n+1}$. 

 This property is the most important assumption made in the  quantitative
 estimates of observable discreteness of background radiation anisotropy  derived below. It is probably general enough to
 apply across a  broad
 class of theories,  including those now receiving the most theoretical attention. Indeed, the main point is that anisotropy
data may provide our first direct view of the underlying,  basic elements or eigenstates of the fundamental theory.

The holographic conjecture has been analyzed in a variety of relativistic  cosmologies,
including de Sitter and anti-de Sitter space  
\cite{susskindwitten,fischler99,bousso,Bousso:1999cb,Bousso:2000dw,bousso99,witten01}.
 The cosmological version of the holographic bound is \cite{Nbound,flanagan} that the
``observable entropy''\footnote{The observable entropy corresponds to the information contained in
a ``causal diamond'', a spacetime volume bounded by two intersecting light cones,
one  open in the future direction of the beginning of an experiment and one in the past
direction of the end of the experiment.} of any universe cannot exceed $S_{max}= 3\pi/\Lambda$,
where
$\Lambda$ is the cosmological constant in Planck units.  In a de Sitter
universe, as in a black hole, this corresponds to  one quarter of the area of
the event horizon in Planck units, but the bound is conjectured to hold for any
spacetime, even Friedmann-Robertson-Walker (FRW) universes with matter as well as 
$\Lambda$.   In the de Sitter case, which we will adopt as a local model for the inflationary
spacetime, 
$S_{max}=\pi/H_{}^2$, where $H$ is the   expansion rate.

 The  
bound includes all degrees of freedom of  all matter fields as well as all  quantum
degrees of freedom of the spacetime itself.
 It implies for example, that 
 a number with $\pi /H_{}^2\ln 2$ bits is sufficient to specify everything that is
going on in a causal diamond of an inflationary universe.
We   consider  the constraints imposed by the entropy bound on     a particular
experiment conducted by nature during inflation: the  formation of classical perturbations from
quantum fluctuations--- or equivalently, nonadiabatic particle production and phase wrapping of a
highly squeezed field state.

\section{Information constraints on field modes   during inflation}

We adopt as an idealized model of the inflationary period a portion of de Sitter
spacetime \cite{Nbound,hawkingandellis}. Figures  1 and 2 illustrate different views of de Sitter spacetime, each showing two different
ways of laying down time and space coordinates. Usually, the field
theory analysis of inflationary perturbations is done with the FRW-like slicing of de Sitter
spacetime, since these space slices  map  directly onto the usual spacelike hypersurfaces of the postinflationary
FRW metric. The metric in this slicing, which covers half of the full de Sitter solution,
appears to have flat spatial slices:
\begin{equation}\label{eqn:metric1}
ds^2=-d\tau^2+\exp[2H\tau]dx^idx_i.
\end{equation}
The comoving modes have fixed lengths in the Euclidean spatial coordinates $x_i$.

The Penrose diagram for a whole inflationary spacetime, including the FRW part at late times,
is shown in figure (\ref{fig: inflationpenrose}). This diagram shows  the route by which information flows
and by which observable patterns are imprinted  on the background radiation by quantum perturbations. For completeness, figure
(\ref{fig: inflationpenrose}) includes not only the causal path for creating microwave background perturbations, but also the
trajectories for gravitons in a possible high-frequency gravitational wave background, such as might be detected directly by
interferometers such as LIGO, LISA or their successors  at frequencies $\approx 10^{\pm 3}$ Hz. 
It
can be produced by quantum graviton fluctuations during inflation,
 and also   classically by mescoscopic mass motions at later times from  e.g. symmetry breaking,
 dimensional reduction, or reheating.  Holographic  discreteness might  
manifest itself in the directly measured classical metric perturbation   of the inflationary waves, but that effect is beyond the scope
of this paper. 

Another   slicing of the de Sitter spacetime (which resembles an inside-out version of a black
hole spacetime in Schwarzschild coordinates) is described by  the metric
\begin{equation}\label{eqn:metric2}
ds^2=-U(r)dt^2 + U^{-1}(r)dr^2+r^2d\Omega_2^2,
\end{equation}
where
\begin{equation}
U(r)=1-{r^2\over r_0^2},
\end{equation}
and $d\Omega_2$ is the (2D) angular interval.
The event horizon stands at radius $r_0=H^{-1}$,   the Bekenstein-Hawking temperature is
$T_{BH}= (2\pi r_0)^{-1}$, and the event horizon area is 
\begin{equation}
A={4\pi\over H^2}={12\pi\over \Lambda}.
\end{equation}
These coordinates   cover the entire ``causal diamond'' accessible to any actual observation by
an observer at the origin, comprising  a finite spatial region (as opposed to the infinite volume
of the FRW slicing)  subject to the holographic entropy bound.  Figure  1 shows a spacetime embedding diagram of this volume with the
two slicings in physical units, and figure 2 compares their Penrose diagrams.

It has  been established that holography is not consistent with independent field modes\cite{bousso02}, and this
can be seen from considering the flow of information in these figures.
A spacelike hypersurface in the coordinates (\ref{eqn:metric2})  extends into the distant, highly redshifted past
and gets very close to the event horizon. The bulk  of the information from a typical small 3-volume near such a surface in the
distant past ended up getting advected out of the horizon.  (The full
information of a  3-volume only survives from the Planck scale to   the horizon scale for a single Planck-sized
 patch in the
very center of the volume at
$r=0$).  
Following one of the  hypersurfaces (\ref{eqn:metric2}) back to where it lies within a Planck
wavelength of the horizon corresponds to a redshift of only $\sqrt{2/H}$ relative to where it intersects the origin (rather than
$\approx H^{-1}$ which takes a horizon-scale mode to the Planck length). Since this is the regime where we expect quantum-gravitational
information mixing among modes,  the vacua of the
  modes will be affected by quantum gravity at a physical mode wavelength much larger than
the Planck scale.    
The ``stretched event horizon'' of the de Sitter space acts like the event horizon of a
black hole,  thermalizing the system by allowing (strong gravitational) interactions between modes
of different wavelength,  so that unitary evolution is not independent for each comoving-wavelength mode;
instead their information is mixed.

Consider   the holographic constraints on  the propagation of
physical influences from above the Planck scale. 
Note that for realistic   inflation models with $H_{}$ well below the Planck scale, the maximum observable
entropy $S_{max}=\pi/H^2$ is always much less than the
3-D de Sitter volume, $4\pi/3H_{}^3$. This means  
 that the holographic bound
  precludes   information from the super-Planck regime from directly propagating to observable
scales:
there   isn't enough information available in the
de Sitter volume to specify the state of   every mode on   the Planck scale at any given time.
Conversely, when any given mode has expanded to
   the de Sitter scale,  much less than one
  bit of information per de Sitter volume survives from ``its own'' Planck epoch. 
Holography allows at most on average only one bit per mode at a scale $\lambda_{hol}\approx
(\lambda_{Planck}^2/H)^{1/3}$; below this scale holography tells us that it is inconsistent to assume
that the states of each mode can be specified independently, because there is not enough
information available to do so. (Incidentally, this scale is about a fermi today, based
on the current estimates of $\Lambda$; however this  has no effect on any local experiment.)  

 While the  effects of
super-Planckian physics have been a subject of debate within the field theory framework
\cite{brandenberger01,martin01,starobinsky01,niemeyer,easther,hui}, these information-counting  arguments suggests that
in these models, {\it any} field theory omits an important constraint of spacetime quantization on the fluctuations even well below
the Planck scale.   Of course,  counting arguments do  not inform us by what mechanism  the holographic constraint
 intervenes.

\section{discreteness of cosmic background anisotropy}

\subsection{Minimal discreteness: spacetimes quantized in discrete steps of $N$}

The holographic constraint clearly imposes some discreteness on the behavior of an inflationary
spacetime.
 At the very least, if  
the classical $H_{}$   adjusts itself in such  a way that the Hilbert space dimension $N$ 
 (as the fundamental quantity \cite{banks2000}) is to  come out to be
an integer,   
$H$ proceeds through a series of discrete steps with 
\begin{equation} \label{eqn:maximal}
H_i=\sqrt{\pi\over\ln N_i}
\end{equation}
where $N_i$ are integers.
For example, nearly exact Planck-scale inflation 
$H=\sqrt{\pi/\ln(23)}=1.000972\dots$ occurs for $N=23$.

In principle, this property is observable.  The pattern
of microwave anisotropy on the sky  on large scales includes  direct images, created by the
Sachs-Wolfe effect,  of the inflaton perturbations in  about 
$\ell_{rec}^2\approx 10^4$ independent de Sitter volumes, where $\ell_{rec}$ is the
angular wavenumber corresponding to the horizon at last  scattering. (Below this scale the
observed anisotropy is strongly affected by plasma oscillations and propagating modes, so the
primordial information is less directly visible).  Thus in this example we can observe
$10^4$ samplings of a process that originally  only yielded 23 distinguishable outcomes, and we
should be able to discern repeated patterns. 

Of course, we cannot yet predict what those patterns are, since we
don't know what  the states of quantum gravity look like (although Bekenstein\cite{bekenstein01}
has   made quantitative  conjectures about quantum states of black hole spacetimes).
The quantum gravitational
discreteness  could manifest itself in a very simple way, such as a discrete component to the 
spectrum of spherical harmonics instead of the continuous gaussian random phase distribution
predicted by the field theory approximation; or in a  more subtle way,  as 
repeated patterns of sky pixel amplitudes. In this  case the discreteness may   be more
conspicuous in the phase information (as opposed to  the power spectrum amplitude) of anisotropy.

The number of distinguishable states grows exponentially, 
$N=\exp[\pi H_{}^{-2}]$. If this corresponds to the number of possible options for the horizon-scale
fluctuations in a de Sitter spacetime,
the   holographic discreteness  is  only observable
if
$H_{}$ lies within a factor of a few of the Planck scale.  Even in principle, the 
largest number of independent inflationary horizons on the sky (observable in principle in    an inflationary
gravitational-wave background) is ``only''
$10^{58}$ (the inverse solid angle of a comoving Planck patch); therefore  if $ \pi/H_{}^2 \ln
(10)
\ge 58$, the discrete system becomes indistinguishable from a continuum. Thus we certainly 
need
$H_{}\ge 0.15$ for an observable effect.

We know that $H$ during inflation is not nearly this large.  Quantum graviton
fluctuations   lead to   classical  tensor-mode perturbations of  amplitude 
$h_{rms}\approx H_{}$.
This leads to a general upper bound (discussed in more detail below) on the Hubble
constant during inflation,  from 
 observations of large angular scale background anisotropy, of about
$H_{}\le 10^{-5}$. The maximum observable de Sitter entropy exceeds
$S_{max}  \approx 10^{10}$, which means that the  total number of quantum states   exceeds
$\exp[10^{10}]$. If the number of
  distinguishable states accessible to an inflaton mode when its
wavefunction collapses were  this large, 
the usual continuum
approximation would   be  a good one for  all   conceivable
microwave background observations on large scales. 

However, this estimate  does not apply in many theories. Holographic discreteness in string theory and loop quantum gravity, like
discreteness in the Hilbert spaces of more familiar quantum systems such as collections of spinning particles, probably derives
ultimately from discrete elements in the fundamental theory. In the remainder of this paper we assume that it is these elements, and
not the states, which can only be added ``one at a time''.  In this case, fluctuations are quantized in discrete steps of $n$ (or $S$)
rather than $N$.

\subsection{Fluctuations quantized in discrete steps of $S$}

Holography tells us that the maximum observable 
total entropy and the de Sitter expansion rate are connected by
$S_{max}=\pi/H^2$. We assume that this connection arises from fundamental discrete elements so that
 $S$ (or $n=S/\ln 2$) occurs  in
 integer steps, fixing a finite set of  discrete values for
$H$. Scalar perturbations, which appear today as observable temperature perturbations,    are determined
classically by
  the dynamics of  the inflaton field $\phi$;   the inflaton perturbations originate as quantum states, and
when they freeze, they in turn 
 directly fix  $H$,   so that their configurations  are also constrained to a finite discrete set of options.
 We now proceed to estimate   observable discreteness--- characterized by the number of options available, or the amount of information
contained in the classical scalar perturbations--- following these connections.

To estimate the   graininess associated with holographic constraints on inflaton fluctuations, we
consider a very simple  toy model, where  observable regions of quasi-de Sitter spaces   occur in a
discrete set of eigenstates of
$H$ with  splitting $\Delta H$ between adjacent  levels: 
\begin{equation}
\dots \leftrightarrow | H_0+ \Delta H> \leftrightarrow | H_0 > \leftrightarrow | H_0- \Delta H>
\leftrightarrow | H_0 -2\Delta H >
\leftrightarrow  |H_0
- 3\Delta H>\leftrightarrow \dots,
\end{equation}
where $| H_0 >$ denotes  a reference de Sitter space of with expansion rate $H_0$. The secular
classical evolution of an inflationary spacetime represents a steady rightward flow, and quantum fluctuations can go
in either direction (although the field-theory fluctuations are in general coherent over many of these
 transitions). This leads to  discreteness in observable anisotropy with steps of some amplitude $\Delta T/T$.

The  key assumption of the toy model is that  {\it the background spacetime  and   horizon-scale
  fluctuations occur through a
sequence of  transitions between discrete   states, each of which adds (or subtracts)
  one bit of information  to the total maximum observable entropy.}
That is, instead of the $H$ coming in discrete steps of $N$ as in  Eq.  (\ref{eqn:maximal}), it comes in discrete
steps of $n$:
\begin{equation} \label{eqn:minimal}
H_i=\sqrt{\pi\over n_i \ln 2}
\end{equation}
where $n_i$ are integers. Once we have assumed discrete steps in $H$,
this  {\it ansatz} represents a plausible lower  bound on the step  size    required for consistency: if $\Delta H$
were smaller, the total entropy increment (in all modes of all fields) would be less than one bit, which is not
enough even to include the degree of freedom represented just by the jump in $H$.

The final
state is the product of  transition operators, the final information the sum of that in the
transitions.
 The change in the dimension of the total Hilbert space for the cumulative effect--- including the change
in the set of all possible field fluctuations--- is still very large, $\delta N=\exp[S_{max}]\delta S_{max}$;  
this   reflects the cumulative effect of choices made in each new bit.  
The holographic bound indicates that  the Hilbert space is the same size as a collection of $n$ spins; in this
model,
 spacetimes  are assumed to  exhibit the same discreteness as  a discrete integer number of such spins.

To estimate the observable  consequences of the toy model,
we use    the classical dynamics of the  inflaton, the bulk of which forms a nearly  
  homogeneous zero-momentum condensate.
The classical expectation value $\phi_c$ of the inflaton
  field has
a unique connection to the entropy through its effect on the spacetime geometry:
the
effective potential
$V(\phi_c)$ and de Sitter   rate $H$ that controls the total entropy are related via the
Friedmann equation,
\begin{equation} \label{eqn:friedmann}
H^2={8\pi \over 3}\left[V(\phi_c)+ {\dot\phi_c^2/2}\right].
\end{equation}
Therefore, large-scale classical perturbations in the inflaton field are directly mapped onto 
total observable entropy and also, in the toy model, come in discrete steps.
 We   assume that
in the discrete sequence of de
Sitter eigenstates,    eigenstates of
$H$ are also eigenstates of
$\phi_c$.

  The  background (unperturbed, continuous, classical)  evolution of
a  homogeneous noninteracting scalar field
$\phi_c$,  is controlled by the classical dynamical equation\cite{lythriotto,pocket}
\begin{equation}
\ddot\phi_c+3H\dot\phi_c+V'(\phi_c)=0,
\end{equation}
where $V'\equiv dV/d\phi_c$. 
We   assume for simplicity that the observed modes crossed the
inflationary horizon  during a standard so-called ``slow roll''  or Hubble-viscosity-limited phase
of inflation, corresponding to $V'/V\le\sqrt{48\pi}$ and $V''/V\le 24\pi$, during which   $\dot\phi_c\approx -V'/3H$.
The rate of the roll is much slower than the expansion rate $H$, so the kinetic term in 
 Eq.  (\ref{eqn:friedmann}) can be ignored in   the mean evolution.
The 
  slow-roll phase of inflation creates    approximately
scale-invariant curvature perturbations. 

We define a
 combination of inflationary parameters   by
\begin{equation}
Q_S\equiv {H^3\over|V'|} \approx \left({V^3\over V'^2}\right)^{1/2} \left({8\pi\over 3}\right)^{3/2}.
\end{equation}
This combination  controls   the Hilbert  space attached to the  perturbations. It also happens  that
it can be estimated
fairly accurately, since it also controls the amplitude of the   scalar perturbations observed in the microwave
background anisotropy. Adopting the notation of
\cite{smootscott}, the anisotropy can be decomposed into scalar and tensor components,
\begin{equation}
{\langle Q \rangle^2\over T_\gamma^2}=S_Q+T_Q,
\end{equation}
where $\langle Q \rangle$ denotes the (global) mean quadrupole anisotropy amplitude,
 $T_\gamma=2.725\pm 0.002$ the mean temperature\cite{smootscott},
and the scalar and tensor contributions  $S_Q$ and $T_Q$ to the mean quadrupole $C_2$  are given by\cite{pocket}
\begin{equation}
S_Q={5C_2^S\over 4\pi}\approx 2.9 {V^3\over V'^2}
\end{equation}
and
\begin{equation}
T_Q={5C_2^T\over 4\pi}\approx 0.56 V,
\end{equation}
using numerical factors  
for a slow-roll, flat $\Lambda$ model. The best fit to the
four-year COBE/DMR data, assuming scale invariance ($n=1$, and $T_Q=0$), yields \cite{bennett,gorski}
$\langle Q \rangle=18\pm 1.6 \mu$K, and hence combining the above,
\begin{equation}
Q_S=9.4\pm 0.84\times 10^{-5}.
\end{equation} 
(Note that if we instead assume $S_Q=0$, these formulae also yield the upper bound $H\le 2\times 10^{-5}$
discussed above, from the tensor modes alone.)

During slow-roll inflation, there is a steady increase in observable entropy at a rate
\begin{equation} \label{eqn:rate}
\dot S_{max}= {8\pi^2\over 9}{V'^2\over H^5}
=8\pi^2 H Q_S^{-2}.
\end{equation}
Every inflationary $e$-folding, $S_{max}$ increases by an amount of order $10^{10}$ due just to the
classical evolution of the system. (In a time $\approx 10^{-10}/H$, the total observable entropy changes by about one
bit.)
The important point, which we will now elaborate, is that the information attached to the observable quantity--- the
horizon-scale perturbation in  the inflaton--- is much smaller than this, of order $10^5$. The basic reason is that
the horizon-scale perturbations contribute only a small fraction $\approx 10^{-5}$ of the total change in observable
entropy; consequently, we conjecture that a typical perturbation can be completely characterized by $\approx 10^5$
bits per Hubble volume. The framework for this argument is illustrated in figure (\ref{fig: slowroll}).

Consider a spatially
uniform change
$\delta\phi_c$ in the classical  inflaton condensate, extending over a volume greater than
$(4\pi/3)H^{-3}$. This
  changes the total observable de Sitter entropy in the affected volume by an amount
\begin{equation}
\delta S_{tot}= {-2\pi \over H^3}\delta H= {-2\pi \over H^3}{4\pi\over 3H} V'\delta \phi_c.
\end{equation} 
Now  consider the effect of a quantum  perturbation in the inflaton field of amplitude
$\delta \phi$, assuming that it behaves like a
``frozen in''   classical spacetime background. 
Substituting from the classical slow-roll formulae above,
an estimate for the   jump in total entropy
associated with a horizon-scale perturbation
$\delta\phi $ can   be directly expressed in terms of    the  observable quantity $Q_S$,
\begin{equation}
\delta S_{tot}={8\pi^2\over 3} \left[{\delta\phi\over H}\right] Q_S^{-1}.
\end{equation}

 The   standard field theory
 analysis   for the horizon-size
perturbations predicts that 
as they are   frozen in  (or squeezed) into a classical state,
the quantity $[\delta \phi/H]$ is   statistically determined, with a   continuous
gaussian statistical distribution of order unit width.
The corresponding increment $\delta S_{qft}\approx   (8\pi^2/3) Q_S^{-1}$ then is roughly the jump in the total
observable cosmological entropy associated with the creation or destruction of a single
horizon-scale inflaton quantum.

{\it  Therefore,   the   standard quantum
transition corresponding to a
 typical perturbation yields a total change of no more  than about $\delta S_{qft}\approx
  (8\pi^2/3)Q_S^{-1}\approx 10^5$ in the maximum observable entropy.}
This is much less than the increase given by Eq. (\ref{eqn:rate}) in the total information 
during an expansion time.
Thus the amount of information $\delta S_{qft}$ attached to the  perturbation in the
horizon-size inflaton quantum $\delta \phi $ as it freezes into a classical  state is much less than the total growth 
of information of the
spacetime during the same time, of order $Q_S^{-1}$ rather than  $Q_S^{-2}$. 
This may be enough to produce a detectable level of discreteness.

Is this story consistent with    the more general 
  ideas that motivated holography in the first place? 
 The naiive assumption in our toy model is that
the dimension of the observable Hilbert space changes  in steps of one bit in the exponent, with each binary choice corresponding to 
a transition to a new quasi- de Sitter eigenstate. This does not mesh trivially with  the idea of preserving overall unitary
evolution so this cannot be quite the whole story; but indeed it is not a trivial matter to even define unitarity
rigorously in 
deSitter spacetime\cite{witten01}.
Another comment is that modes much smaller than the
horizon have their accessible Hilbert space changed due to a quantum fluctuation on the horizon scale.
From the point
of view of an observer at rest in the center of a ``$\delta
\phi$ patch,''   much of the incremental information attached to
$\delta
\phi $ appears in modes much smaller than
$H^{-1}$--- the short-wavelength modes close to the stretched event horizon that are still
having their vacuum initial states formed, and only later expand to the horizon scale. Much
of the ``new'' information represented by $\delta S_{tot}$ must be available to modes  near the
stretched event horizon as viewed by any given observer--- modes which are much smaller than
$H^{-1}$. Thus it makes sense that the information available to  the configurations of horizon-scale
fluctuations  is much smaller than
$S_{max}$.

\subsection{Toy Model of Anisotropy with Discrete Levels of $T$}

It is useful to delve into some more detail with a more concrete, albeit  unrealistic,
toy model for discreteness in the perturbations.
Suppose that the possible values of the horizon-size inflaton perturbation
$\delta\phi$ are     selected from a discrete set of levels
 separated by some step size $\Delta \phi$. The   
incremental observable information   made available by the transition
between levels is
\begin{equation} \label{eqn:totalentropychange}
\Delta S_{tot}={8\pi^2\over 3} \left[{\Delta\phi\over H}\right] Q_S^{-1},
\end{equation}
including  all the degrees of freedom of all the fields,
not just the inflaton. 
We now assert that {\it  jumps in $\delta\phi$    occur 
 in steps of at
least a certain minimum size $\Delta\phi$, such that $\Delta S_{tot}\ge\ln 2$ 
(that is, a change of at least one bit in total observable entropy)}. This leads to a 
definite minimum step size,
\begin{equation}
 \left[{\Delta\phi\over H}\right]  \ge  Q_S (3 \ln 2/8\pi^2)= 2.5\times 10^{-6}.
\end{equation}
As above, this {\it ansatz} seems necessary for consistency: a step at least this large is 
needed just to specify the degree of freedom represented by the $\Delta\phi$ transition itself.

Taking this model literally and assuming subsequent linear classical evolution,
this discreteness   of the field
amplitudes leads to a discrete set of values for the curvature perturbation (e.g., Bardeen's\cite{bardeen1980}
$\phi_m$), and therefore the observed  temperature perturbation $\delta T$, which is largely determined by this
quantity via the Sachs-Wolfe effect. 
 The discrete jumps are   smaller than the total standard predicted  perturbation amplitude by
about the  factor
\begin{equation}
{\Delta T\over \delta T}\approx {\Delta \phi\over \delta \phi}.
\end{equation}
 That is, the  amplitude  of the anisotropy on the sky,  instead of coming from
a continuous distribution (with variance $\approx \langle Q \rangle^2$ contributed by fluctuations
in each octave of angular wavenumber),
comes from a discrete distribution,   with     values of $T$
formed by a sum of discrete increments  of order $\Delta T$, where
\begin{equation} \label{eqn:graininess}
\left[{ \Delta T \over  \langle Q \rangle}\right]\approx 
 \left[{\Delta\phi\over H }\right].
\end{equation}
The  amplitude difference  at which the discreteness appears is about
\begin{equation}
\Delta T \ge \left[{\Delta\phi\over H}\right] \langle Q \rangle\approx   10^{-10}\ {\rm K}.
\end{equation}

\section{Observations?}

The toy-model results      should not be taken as a literal physical prediction of
the character of the discreteness; such a prediction would require understanding of the true
eigenmodes of the system.  We do not know from these arguments what their
detailed character is, or how they might be manifested in the  $A_{\ell m}$ spectrum of the anisotropy.  
(For this reason, we have also glossed over the facts that observed anisotropy is a superposition  of modes from
different comoving scales, and that these are increasingly affected, on scales smaller than the quadrupole, by
propagation effects before,  during and after recombination that modify  the  primordial signal\cite{bond01}).

The analysis above does  however highlight  several  new and unexpected features that might apply to the real-world
discreteness: (1) The characteristic fractional amplitude of the inflaton perturbation discreteness  is not
exponentially small, but is comparable to the overall scalar perturbation
amplitude. (2)   The information in a typical horizon-scale perturbation does not depend explicitly on unknown
parameters such as the   values of
$H$ or $V'$, except through the observable combination $Q_S$. 
(3) {\it The   holographic counting arguments   suggest    that an inflaton perturbation at the
time it freezes out  has a Hilbert space   equivalent to only about $ 10^5$ binary spins.} 
We therefore  conjecture   that the
theory incorporating the true eigenmodes may predict  some (unspecified) kind of   discreteness corresponding to 
less than about $  10^{5}$ bits of information per mode.

The observability of discreteness depends  greatly on how 
it is manifested.
To choose a somewhat silly  example, suppose that rather than
occuring in $\approx 10^5$ discrete levels of amplitude, a  perturbation is composed of $\approx 10^5$ 
discrete  binary pixels on the sky. This level of pixelation in the quadrupole modes (say) 
 would not be completely erased by transport effects at recombination, which propagate on an angular scale comparable
to this pixel size (for example, the angular wavenumber of the first ``acoustic peak'' in the angular power spectrum
$C_\ell^2$  is  at $\ell\approx 200$, corresponding to $\ell^2\approx 4\times 10^4$ pixels). 
With discreteness encoded this way, each pixel might be either hot or cold, $\delta T/T=\pm 1 \times 10^{-5}$ (say).
Intermediate gray levels would appear as a checkerboard pattern of alternating hot and cold pixels
on a scale of about
a degree of arc. Such a pattern would be  very conspicuous in the current data,
 even allowing for the complication that it would be superimposed on
approximately gaussian noise from smaller scale modes of comparable overall amplitude.

Thus, for some kinds of eigenmodes,  observations of  holographic discretness may be practical.
 The   possibility of finding
such a qualitatively new effect motivates a search in the microwave background data for  discrete behavior, in either
the amplitudes or the multipole phases.   
Some tests are straightforward; for example, a histogram of $A_{\ell m}$ amplitudes normalized to a best-fit smooth
$C_{\ell}$ spectrum might show  statistically significant departures from a gaussian distribution. 
  These discreteness effects are qualititively different from
(and have a more distinctively
quantum character than)  other possible
new effects, for  example  
string-theory-inspired predictions\cite{easther,kaloper} of departures  from  standard inflationary theory  for the
tensor and scalar power spectra. 
However, if the level of discreteness is
as small as  
$\Delta T/T\approx  10^{-10}$, and appears only as  discrete levels of 
Sachs-Wolfe perturbation amplitude (say),  plasma motions   during recombination, as
well as nonlinear couplings, are likely to smear out all the   observable traces of primordial discreteness even on
the largest scales.

\acknowledgements

I am grateful for useful conversations with J. Bardeen, R. Bousso, and R. Brandenberger.
 This work was supported by NSF grant AST-0098557 at the University of Washington.


{}
\break

\begin{figure}[htbp]
\centerline{\epsfig{height=5in,file=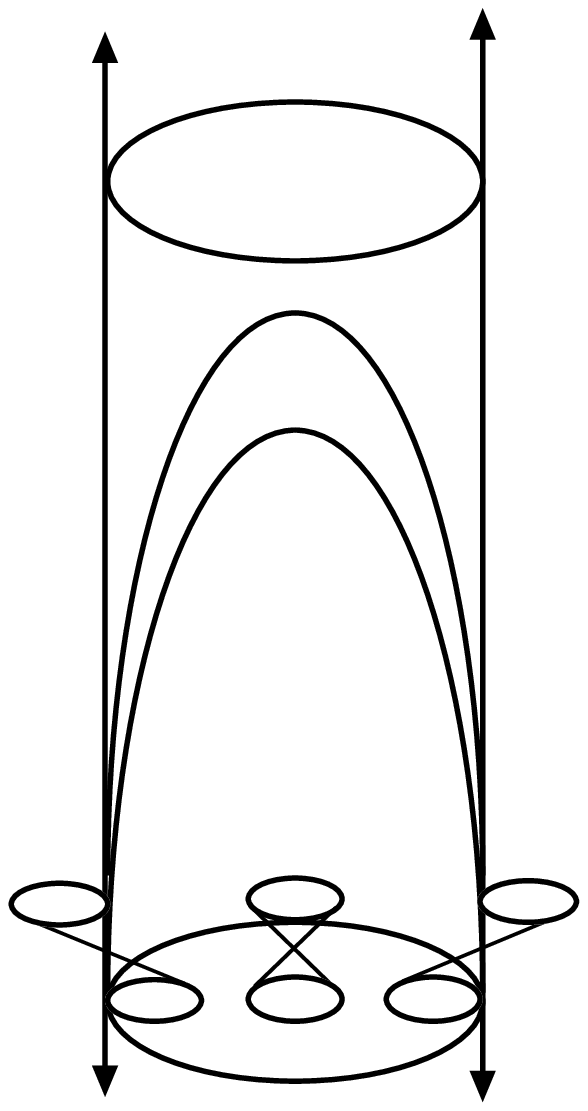}}
\caption{ \label{fig: desitter} Embedding diagram of an observable region of de Sitter
spacetime, with one spatial dimension supressed, in proper physical units. The cylinder represents the world-sheet swept
out by the  event horizon (a 2-sphere) of an inertial observer at the center. Horizontal sheets
correspond to the  slices of constant time  in  Eq.  (\ref{eqn:metric1}); curved sheets
represent schematically the behavior of the constant-time surfaces in Eq. (\ref{eqn:metric2}).}
\end{figure}

\begin{figure}[htbp]
\centerline{\epsfig{height=6in,file=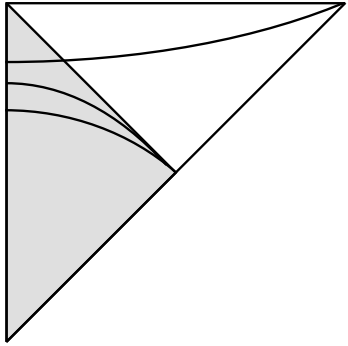}}
\caption{\label{fig: penrose} Penrose diagram of the portions of de Sitter space described by the
metrics,
Eq. (\ref{eqn:metric1}) and Eq. (\ref{eqn:metric2}), showing the same spatial slicings as Fig. (1).  The
shaded region corresponds to the observable region shown in Fig. (1) (the whole region described
by Eq. (\ref{eqn:metric2})). }
\end{figure}

\begin{figure}[htbp]
\centerline{\epsfig{height=6in,file=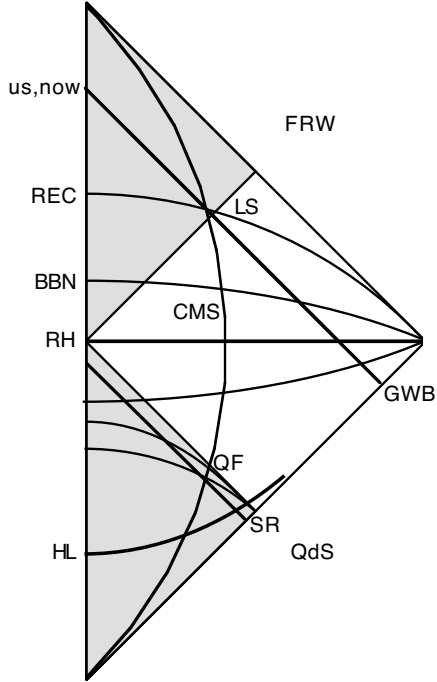}}
 \vspace{0.5in}
\caption{\label{fig: inflationpenrose} 
Penrose diagram of  an inflationary cosmology, showing the route of  information flow from inflation to observable anisotropy. As
usual, points represents spheres, the left hand edge represents the world
line of an observer at the origin, and  the extremities represent
boundaries at infinity. The lower half  represents a quasi- de Sitter
space (QdS), the upper half represents a
 Friedmann-Robertson-Walker space (FRW).
 The join between them is the
epoch of reheating (RH), and shaded regions of each show the regions with
the ``apparent horizon'' of an observer at the origin  (apparent event
horizon for QdS, apparent particle horizon for FRW).  Labled spacelike
hypersurfaces include  the recombination epoch (REC) and Big Bang
Nucleosynthesis (BBN). Spacelike hypersurfaces in the QdS phase are shown
for two different slicings, one appropriate for matching onto FRW and the
other for holographic analysis. The intersection of our past light cone
with the recombination surface is the two-sphere of the ``last scattering
surface'' (LS) of the cosmic background radiation.
The timelike trajectory of a
comoving sphere (CMS)  is shown, first within the inflationary event
horizon, then passing outside of it, then eventually reentering the
apparent particle horizon during the FRW phase. The particular CMS shown
is one that enters close to recombination, and therefore is on a scale that affects the
observed microwave anisotropy. Perturbations are imprinted by
fluctuating quantum fields (QF) on the scale of the apparent horizon
during  the slow-roll period of inflation (SR).
(The apparent
horizon grows slightly during SR as indicated by the two closely
parallel null lines, as expanded in Fig. (\ref{fig: slowroll}).) This includes
contributions from both tensor and scalar modes that are  frozen in  to
the classical metric outside the apparent horizon. 
 A high-frequency gravitational wave background (GWB) reaches us 
  via direct null trajectories. 
The classical continuum picture breaks down for all past trajectories beyond some holographic limit (HL).}
\end{figure}

\begin{figure}[htbp]
\centerline{\epsfig{height=4in,file=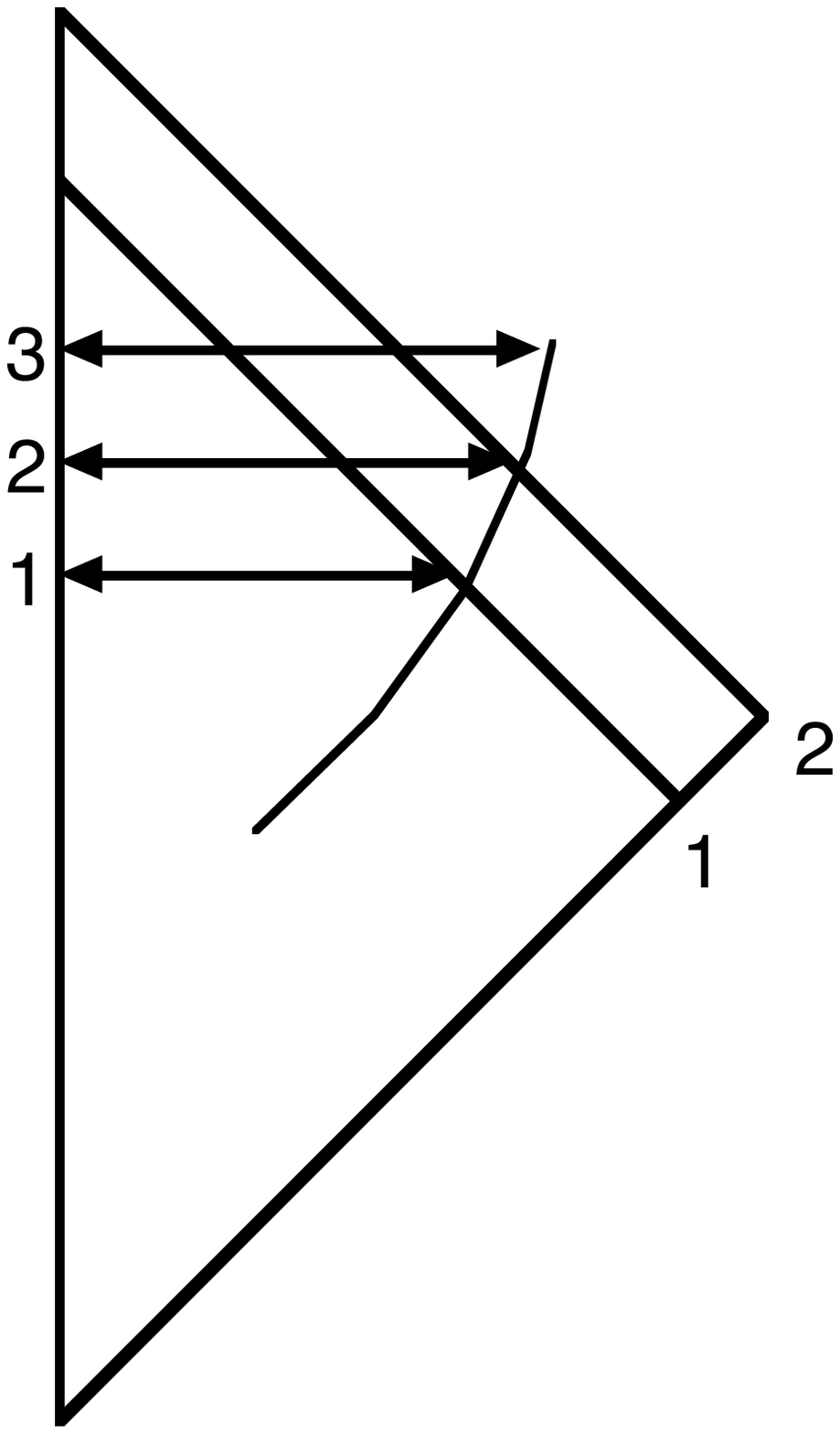}}
 \vspace{1in}
\caption{\label{fig: slowroll} A close-up view of  freeze-out:
the formation and  ``collapse'' of a quantum fluctuation into a classical
perturbation.  It takes about one Hubble time
for a perturbation on the horizon scale to  freeze out. During this time,
the inflaton rolls slightly and
$H$ decreases slightly; the corresponding   apparent horizons are shown as 1
and 2 in the figure. Arrows indicate the physical size of a comoving mode as
it passes through the horizon. By time 3, the perturbation is ``outside the
horizon'' and the  frozen-in perturbed value of the inflaton fixes the expansion
rate of the background  spacetime on subhorizon scales. 
 The total increase in entropy during the interval (1,2)
(that is, the difference in entropy between the two triangles) is 
about $10^{10}$, but the frozen-in  perturbation represents only a small
fraction of these degrees of freedom, about $10^5$.}
\end{figure}


\begin{thebibliography}{}

\bibitem{Starobinsky:1979ty}
A.~A.~Starobinsky,
JETP Lett.\  { 30}, 682 (1979)
[Pisma Zh.\ Eksp.\ Teor.\ Fiz.\  { 30}, 719 (1979)].


\bibitem{Hawking:1982cz}
S.~W.~Hawking,
Phys.\ Lett.\ B { 115}, 295 (1982).

\bibitem{Guth:1982ec}
A.~H.~Guth \& S.~Y.~Pi,
Phys.\ Rev.\ Lett.\  { 49}, 1110 (1982).

\bibitem{Bardeen:1983qw}
J.~M.~Bardeen, P.~J.~Steinhardt \& M.~S.~Turner,
Phys.\ Rev.\ D { 28}, 679 (1983).

\bibitem{Starobinsky:1982ee}
A.~A.~Starobinsky,
Phys.\ Lett.\ B { 117}, 175 (1982).

\bibitem{Halliwell:1985eu}
J.~J.~Halliwell \& S.~W.~Hawking,
Phys.\ Rev.\ D { 31}, 1777 (1985).

\bibitem{Grishchuk:1993ds}
L.~P.~Grishchuk,
Class.\ Quant.\ Grav.\  { 10}, 2449 (1993)


\bibitem{bardeen1980}
J. M. Bardeen, Phys Rev D 22, 1882 (1980)

\bibitem{cobeDMR}
 G.F.Smoot et al.,Astrophys.J. 396, L1(1992) 

\bibitem{bennett}
C. L. Bennett et al., Astrophys. J. 464, L1 (1996)

\bibitem{gorski}

K. M. Gorski et al., Astrophys. J. 464, L11 (1996)



\bibitem{boom}
 P. de Bernardis et al.,
Nature 404 (2000) 955-959

\bibitem{pryke}
C. Pryke et al.,
Ap. J. in press, astro-ph/0104490 (2001)
 
 
\bibitem{halverson}
N. W. Halverson et al., 
Angular
 astro-ph/0104489 (2001)


\bibitem{max} S. Hanany et al.,
 Astrophys.J. 545, L5  (2000)



\bibitem{bond01}
J.  R. Bond  \&  R. G. Crittenden,
``CMB Analysis'', in Proc. NATO ASI ``Structure Formation in the Universe'', eds. R.G. Crittenden \& N.G. Turok,
astro-ph/0108204,
(Kluwer, 2001)







\bibitem{Grishchuk:1989ss}
L.~P.~Grishchuk \& Y.~V.~Sidorov,
Class.\ Quant.\ Grav.\  { 6}, L161 (1989).

\bibitem{Grishchuk:1990bj}
L.~P.~Grishchuk \& Y.~V.~Sidorov,
Phys.\ Rev.\ D { 42}, 3413 (1990).


\bibitem{Albrecht:1994kf}
A.~Albrecht, P.~Ferreira, M.~Joyce \& T.~Prokopec,
Phys.\ Rev.\ D { 50}, 4807 (1994)

\bibitem{Lesgourgues:1997jc}
J.~Lesgourgues, D.~Polarski \& A.~A.~Starobinsky,
Nucl.\ Phys.\ B { 497}, 479 (1997)



\bibitem{Polarski:1996jg}
D.~Polarski \& A.~A.~Starobinsky,
Class.\ Quant.\ Grav.\  { 13}, 377 (1996)

\bibitem{kiefer}
  C. Kiefer, D. Polarski \& A.A. Starobinsky,
Int.J.Mod.Phys. D7,
 455-462 (1998)





\bibitem{Mukhanov:1992me}
V.~F.~Mukhanov, H.~A.~Feldman \& R.~H.~Brandenberger,
Phys.\ Rept.\  { 215}, 203 (1992) 

\bibitem{lythriotto}
 D. H. Lyth \& A. Riotto, 
Phys.Rept. 314,  1-146 (1999)


\bibitem{Bekenstein:1972tm}
J.~D.~Bekenstein,
Lett.\ Nuovo Cim.\  { 4}, 737 (1972) 

\bibitem{Bardeen:gs}
J.~M.~Bardeen, B.~Carter and S.~W.~Hawking,
Commun.\ Math.\ Phys.\  {\bf 31}, 161 (1973).





\bibitem{Bekenstein:1973ur}
J.~D.~Bekenstein,
Phys.\ Rev.\ D { 7}, 2333 (1973) 

\bibitem{Bekenstein:1974ax}
J.~D.~Bekenstein,
Phys.\ Rev.\ D { 9}, 3292 (1974)



\bibitem{Hawking:1975sw}
S.~W.~Hawking,
Commun.\ Math.\ Phys.\  { 43}, 199 (1975) 


\bibitem{Hawking:1976ra}
S.~W.~Hawking,
Phys.\ Rev.\ D { 14}, 2460 (1976) 



\bibitem{Bekenstein:1981jp}
J.~D.~Bekenstein,
Phys.\ Rev.\ D { 23}, 287 (1981) 



\bibitem{bekenstein02}
J. D. Bekenstein \& G. Goury,
``Building blocks of a black hole'',
gr-qc/0202034 



\bibitem{Gibbons:1977mu}
G.~W.~Gibbons \& S.~W.~Hawking,
Phys.\ Rev.\ D { 15}, 2738 (1977) 

\bibitem{Banks:1984by}
T.~Banks, L.~Susskind \& M.~E.~Peskin,
Nucl.\ Phys.\ B { 244}, 125 (1984) 



\bibitem{'tHooft:1985re}
G.~'t Hooft,
Nucl.\ Phys.\ B { 256}, 727 (1985) 


\bibitem{Susskind:1993if}
L.~Susskind, L.~Thorlacius \& J.~Uglum,
Phys.\ Rev.\ D { 48}, 3743 (1993)


\bibitem{stephens}
 C.R. Stephens, G. 't Hooft, \& B.F. Whiting,
 Class. Quant. Grav. 11, 621-648  (1994)


\bibitem{thooft93}
G. 't Hooft,
`` Dimensional Reduction in Quantum Gravity'',
essay in honor of Abdus Salam, gr-qc/9310026 (1993)

\bibitem{susskind95}

L. Susskind,
J.Math.Phys. 36, 6377   (1995)


\bibitem{Bigatti:1999dp}
D.~Bigatti and L.~Susskind,
 ``TASI lectures on the holographic principle,''
arXiv:hep-th/0002044 (1999)

\bibitem{bekenstein01}
J. D. Bekenstein, ``Quantum Information and Quantum Black Holes'',
Lectures delivered at  NATO Advanced
Study Institute: International School of Cosmology and Gravitation: 17th Course
of the International School of Cosmology and Gravitation, Erice, May 1-11, 2001,   to appear in
Advances in the Interplay between Quantum and Gravity Physics, ed. V. de Sabbata (Kluwer,
Dordrecht 2002)   


  
\bibitem{bousso02}
R. Bousso, ``The Holographic Principle'',
Rev. Mod. Phys., in press (2002),
hep-th/0203101

\bibitem{ashtekar02}
A. Ashtekar,
``Quantum Geometry In Action: Big Bang and Black Holes'',
math-ph/0202008
 

\bibitem{susskindwitten}
L. Susskind \& E. Witten,
``The Holographic Bound in Anti-de Sitter Space'', hep-th/9805114 (1998) 

\bibitem{fischler99}
W. Fischler \& L. Susskind,
``Holography and Cosmology'', hep-th/9806039 (1998)

\bibitem{bousso}
   R. Bousso, 
JHEP 9907, 004 (1999)
 

\bibitem{Bousso:1999cb}
R.~Bousso,
JHEP { 9906}, 028 (1999)


\bibitem{Bousso:2000dw}
R.~Bousso,
Class.\ Quant.\ Grav.\  { 17}, 997 (2000)


 \bibitem{bousso99}
 R. Bousso, 
  Phys.Rev. D60, 063503 (1999)

\bibitem{witten01}
E. Witten, ``Quantum Gravity In De Sitter Space'',
hep-th/0106109

 \bibitem{Nbound}
 R. Bousso, ``Positive Vacuum Energy and the N-bound'', JHEP in press,
hep-th/0010252 (2001)



\bibitem{flanagan}
 E. E. Flanagan, D. Marolf, R. M. Wald,
 Phys.Rev. D62, 084035  (2000)

\bibitem{hawkingandellis}

 S. W. Hawking  \& G. F. R. Ellis, ``The Large Scale Structure of Space-Time,''
Cambridge, 1973


\bibitem{brandenberger01}
 R. H. Brandenberger \& J. Martin,
  Mod.Phys.Lett. A16 (2001) 999 

\bibitem{martin01}
J. Martin \& R. H. Brandenberger,
  Phys.Rev. D63 (2001) 123501

\bibitem{starobinsky01}
A.A. Starobinsky,
 Pisma Zh.Eksp.Teor.Fiz. 73, 415   (2001); JETP Lett. 73, 371   (2001)

 

\bibitem{niemeyer}
J.C. Niemeyer,
  Phys.Rev. D63, 123502  (2001)


\bibitem{easther}
 R. Easther, B. R. Greene, W. H. Kinney \& G. Shiu,
``Inflation as a Probe of Short Distance Physics'', (2001)
hep- th/ 0104102 

\bibitem{hui}
L. Hui \& W. H. Kinney, ``Short Distance Physics and  the Consistency Relation for the Scalar and Tensor Fluctuations
in the Inflationary Universe'',  Phys. Rev. Lett. in press (2001), astro-ph/0109107


\bibitem{banks2000}
T. Banks, ``Cosmological breaking of supersymmetry, or little Lambda goes back to the
future II'' (2000), hep-th/0007146

\bibitem{pocket}
E. W.  Kolb \& M. S. Turner, Eur. J. Phys. C 15, 125 (2000)

\bibitem{smootscott}
G. F. Smoot \& D. Scott, Eur. J. Phys. C 15, 145 (2000)




\bibitem{kaloper}
N. Kaloper, M. Kleban, A. Lawrence, \& S. Shenker,
``Signatures of Short Distance Physics in the
Cosmic Microwave Background'',SU-ITP-02/02, hep-th/0201158

\end{thebibliography}
\end{document}